\documentclass{ws-ijmpe}

\begin{document}

\markboth{V.Kumar and P.C. Srivastava}{Structure of odd $^{79,81,83}$Se isotopes}

\catchline{}{}{}{}{}
%%%%%%%%%%%%%%%%%%%%%%%%%%%%%%%%%%%%%%%%%%%%%%%%%%%%%%%%%%%%%%%%%%%%

\title{STRUCTURE OF ODD $^{79,81,83}$Se ISOTOPES WITH PROTON AND NEUTRON EXCITATIONS ACROSS $Z=28$ AND $N= 40$}

\author{VIKAS KUMAR and P.C. SRIVASTAVA\footnote {pcsrifph@iitr.ac.in}}

\address{Department of Physics, Indian Institute of Technology Roorkee 247 667, India}

\maketitle

%\pub{Received (Day Month Year)}{Revised (Day Month Year)}

\begin{abstract}

The recently measured experimental data of $^{79,81,83}$Se isotopes have been 
interpreted in terms of shell model calculations. 
The calculations have been performed in ${f_{5/2}pg_{9/2}}$ space with the recently
derived interactions, namely with JUN45 and jj44b.  
To study the importance of the proton excitations 
across the $Z=28$ shell in this region. We have also performed calculation in 
${fpg_{9/2}}$ valence space using an ${fpg}$ effective interaction with $^{48}$Ca core and
imposing a truncation. Excitation energies, $B(2)$ values, quadrupole moments 
and magnetic moments are compared with experimental data when available.
Present study reveals the importance of proton excitations 
across the $Z=28$  shell for predicting quadrupole and magnetic moments.

\keywords{monopole; collectivity.}
\end{abstract}

\ccode{PACS Nos.: 21.60.Cs, 27.50.+e.}

%=================================================================
\section{Introduction}
\label{s_intro}
The development of collectivity, island of inversion and single-particle versus collective phenomena
in $40 \leq N \leq 50$ region is the topic of current research for the investigation. 
Experimental evidence of quadrupole collectivity in the neutron rich Fe and Cr with $N \sim 40$ is recently
reported by Crawford et al.\cite{Crawford13} In the novel theoretical work of Zuker et al it was mentioned that
the enhanced quadrupole collectivity in this region due to presence of $0g_{9/2}$ and its quasi-$SU(3)$ 
counterpart $1d_{5/2}$ orbital.\cite{zuker} The interaction for this space recently proposed by Madrid-Strasbourg
group. \cite{Caurier,Lenzi}  Also the importance of  the inclusion of intruder orbitals from $sdg$
shell in the model space for $fp$ shell nuclei is reported in the literature.\cite{Kaneko08,Srie,Sriaa,Sun12,Sun13,Stepp13}
The similarity between island of 
inversion around $N=20$ and collectivity around $N=40$ in Mn isotopes from $^{63}$Mn onwards is
recently reported in.\cite{prc86.014325.2012}
The evolution of collectivity in Ge isotopes with $B(E2)$ measurements have been reported in. \cite{Rodal05}
In this work it is shown 
that the $N=40$ shell closure is collapsed in $^{72}$Ge. However, the $N=50$ shell closure is persistent
in Ge isotopes.  Using the intermediate-energy Coulomb excitation collectivity at $N=50$
for the $^{82}$Ge and $^{84}$Se have been established.\cite{Gade10}

\begin{figure}
\begin{center}
\includegraphics[scale=0.6]{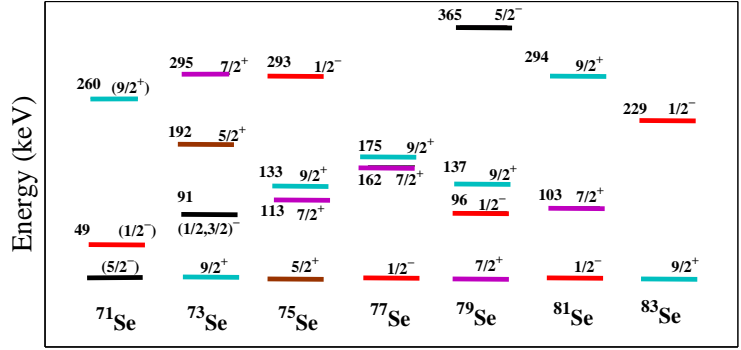}
\end{center}
\caption{
Experimentally observed systematics of low-lying states in odd-A selenium isotopes covering $N=40$ to $N=50$ shell gaps.}

\end{figure}\label{f_seintro}

The low-energy systematics of odd-A selenium isotopes and the evolution of the $1/2^-$, $5/2^-$ and $9/2^+$
levels is shown in Figure 1. It is visible from the figure that in
$^{77,81}$Se isotopes, the ground state is marked
by the $1/2^-$. In case of $^{73}$Se the ground state is $9/2^+$ and it start increasing up to $^{77}$Se, and finally
it becomes ground state for $^{83}$Se.

In the present paper we consider neutron-rich odd Se isotopes. The shell model calculation 
in $f_{5/2}pg_{9/2}$ space for Se isotopes is reported in the
literature using pairing 
plus quadrupole-quadrupole interactions.\cite{yosinaga} 
The importance of inclusion of proton $f_{7/2}$ orbital was pointed out by Cheal et al to explain
sudden structural changes between $N=40$ and $N=50$ for Ga isotopes.\cite{Cheal10}
In our recent investigation~\cite{pcs_ga} we successfully explained
electromagnetic moments of Ga isotopes by including $f_{7/2}$ orbital in
the $f_{5/2}pg_{9/2}$ model space.

The paper is organized as follows. In Section~2 gives details of the shell model (SM) calculations. We will discuss in this section the  model space and
the effective interactions used in the investigation.
Section~3 includes results on the spectra of $^{79,81,83}$Se isotopes  and configuration mixing in these nuclei.
In Section~4, SM calculations on $E2$ transition probabilities, quadrupole moments and magnetic moments are presented. 
Finally, concluding remarks are given in Section~5. 

\section {\label{sec2}Details of Model Spaces and Interactions}

We have performed calculations in two different shell-model spaces.
In case of $f_{5/2}pg_{9/2}$ space we employed two recently derived effective shell model interactions, JUN45
and jj44b, that have been proposed for the 1$p_{3/2}$,
0$f_{5/2}$, 1$p_{1/2}$ and 0$g_{9/2}$ single-particle
orbits.  The JUN45, developed by  Honma {\it et al.} \cite{Honma09}, is a realistic interaction based on the
Bonn-C potential fitting by 400 experimental binding and excitation energy data with mass
numbers $A = $ 63--96.  Brown and Lisetskiy \cite{brown} developed jj44b interaction by fitting 600 binding
energies and excitation energies with $Z = $ 28--30
and $N = $ 48--50. 
The single-particle energies for the 1$p_{3/2}$,
0$f_{5/2}$, 1$p_{1/2}$ and 0$g_{9/2}$ single-particle
orbits employed
in conjunction with the JUN45 interaction are -9.8280, -8.7087, -7.8388, and -6.2617 MeV
respectively. In the case of the jj44b interaction they are -9.6566, -9.2859, -8.2695, and -5.8944
MeV, respectively. The core is $^{56}$Ni, i.e. $N = Z = 28$, and the calculations are performed in
this valence space without truncation.
For the JUN45 and jj44b interactions the single-particle energies are based on those of $^{57}$Ni.
For JUN45 and jj44b interactions there is a rapid decrease
in $f_{5/2}$ proton single-particle energy relative to
$p_{3/2}$ as the neutrons start filling in $g_{9/2}$ orbit and
it become lower than $p_{3/2}$  for $N>48$ this is shown in Figure 2.
\begin{figure}[h]
%\setcaptionmargin{5mm}
%\onelinecaptionsfalse
\resizebox{0.9\textwidth}{!}{
\includegraphics{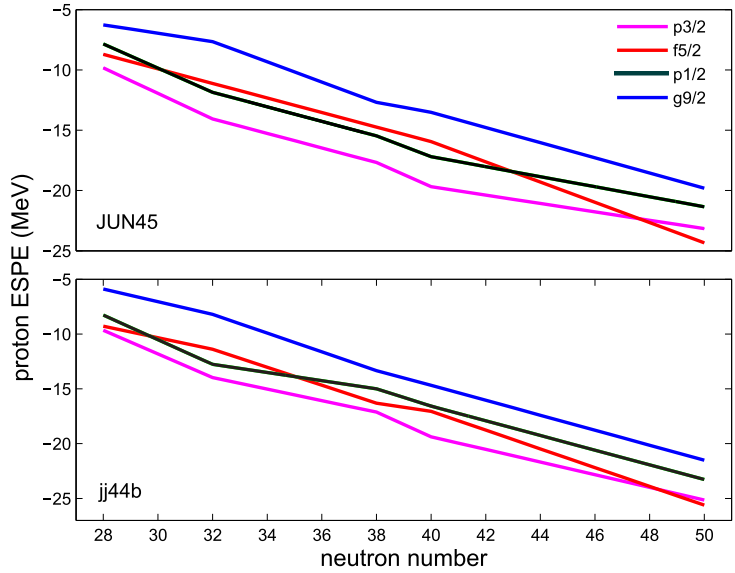} 
}
\vspace{5mm}
\caption{
Effective proton single-particle energies for Cu isotopes for JUN45 and jj44b interactions.}
\label{f_monopole}
\end{figure} 

In the  $f \,p \,g_{9/2}$ valence space,
we use a  $^{48}$Ca core, where eight neutrons are frozen in the $\nu f_{7/2}$
orbital. This interaction was reported  by Sorlin {\it et al}.
\cite{Sorlin02} For this model space we allowed up to a total of four particle
excitations from the $f_{7/2}$ orbital to the upper $fp$ orbitals for protons and from the upper $fp$
orbitals to the $g_{9/2}$ orbital for neutrons.
The $fpg$ interaction for $f \,p \,g_{9/2}$ valence space was built
using $fp$ two-body matrix elements (TBME) from
 \cite{Pov01} and $rg$ TBME ($p_{3/2}$, $f_{5/2}$, $p_{1/2}$ and $g_{9/2}$ orbits) from.\cite{Nowacki96} For
the common active orbitals in these subspaces, matrix elements
were taken from.\cite{Nowacki96} The
remaining $f_{7/2} g_{9/2}$  TBME are taken from.\cite{Kahana69}
The single-particle energies
are 0.0, 2.0, 4.0, 6.5 and 9.0 MeV for the 0$f_{7/2}$, 1$p_{3/2}$, 1$p_{1/2}$, 0$f_{5/2}$, and 0$g_{9/2}$ orbits, respectively.

All calculations in the present
paper are carried out at \textsc{dgctic-unam} computational facility KanBalam
using the shell model
code \textsc{antoine}.\cite{Antoine}
%In case of $^{79}$Se for positive parity
%maximal dimension is 59 791822. For this nuclei the computing time was $\sim$ 12 days for
%both parity.

\section{\label{sec3}Spectra}
 
Shell model results for different model spaces presented in Figures 3-5. Yosinaga et al
\cite{yosinaga} previously presented shell model results in ${f_{5/2}pg_{9/2}}$ space
for pairing plus quadrupole-quadrupole interaction for odd Se isotopes. Present work will
add more information by including $f_{7/2}$ orbital in the model space to study importance
of proton excitation across the $Z=28$ shell. Recently we have reported results for
even Se isotopes in Ref. \cite{pcs_physica} The comprehensive comparison of shell model
results for three interactions used in the calculations are presented with respect to
the experimental data.
    
\subsection{$^{79}$Se}
 
\begin{figure}
%\setcaptionmargin{5mm}
%\onelinecaptionsfalse
%\resizebox{1.03\textwidth}{!}{
%\includegraphics{79Se.eps} 
%}
\includegraphics[width=14.4cm]{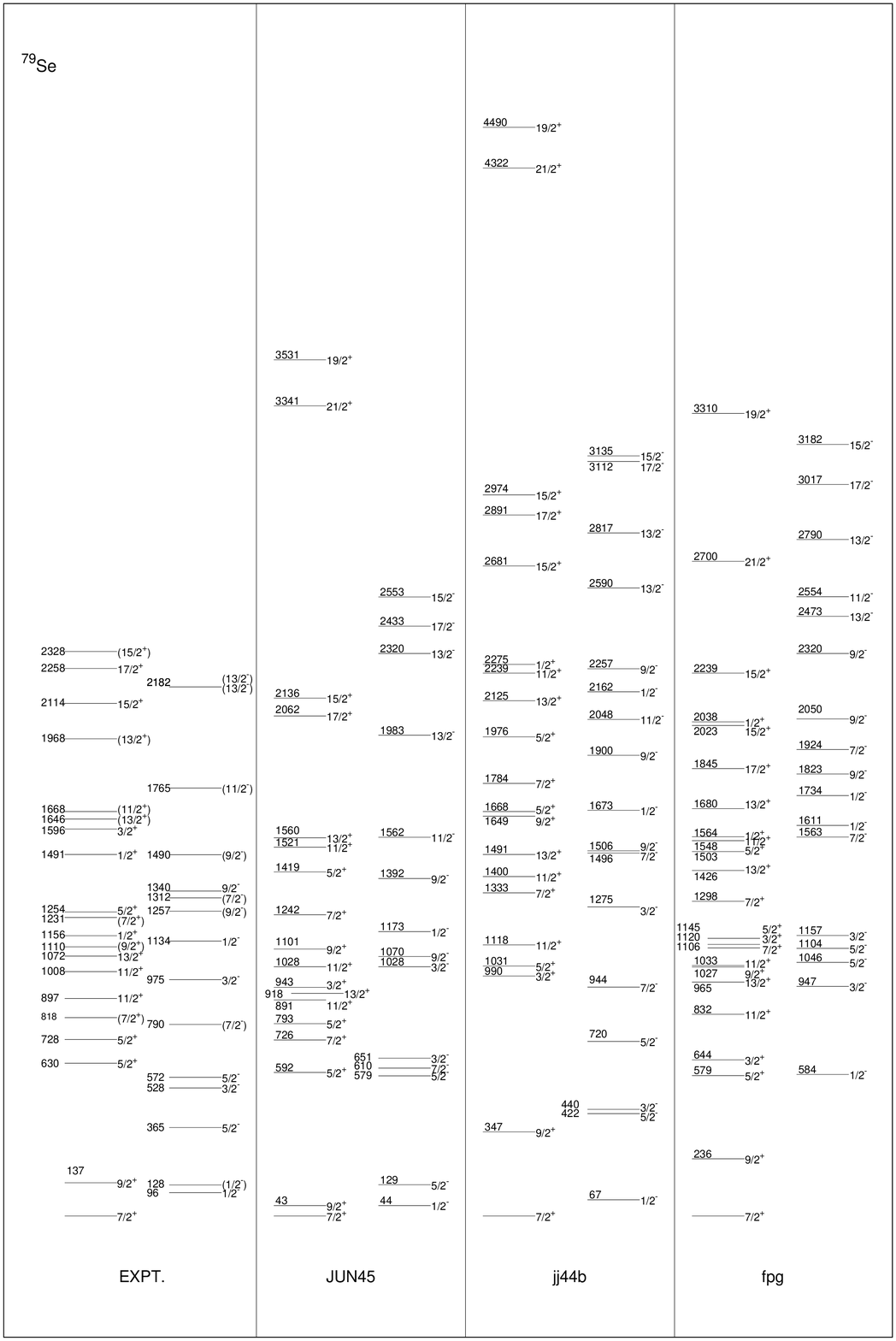}
\caption{
Experimental data  for $^{79}$Se [24] compared
with the results of large-scale shell-model calculations using 
three different effective interactions}
\label{f_se79}
\end{figure}

 The calculated values of the energy levels of $^{79}$Se with the help
 of JUN45, jj44b and $fpg$ interactions are shown in Figure~\ref{f_se79}. 
 All the three interactions correctly reproduced the ground state spin and parity.
 In case of JUN45, the calculated $9/2^+$, $5/2^+$ lower in energy. The
 jj44b interaction predicted $9/2^+$ about 200 keV higher in energy. The second $1/2^-$
 predicted by all interactions are higher in energy while in experiment they are very
 close to each other. The positive parity levels predicted by jj44b interaction is
 higher in energy. All the interaction predict correctly the first negative parity
 as $1/2^-$, but it is lowered by 52 keV in JUN45 and by 29 keV in jj44b.
  The experimental sequence of $5/2^-$, $3/2^-$, $5/2^-$, $7/2^-$, $3/2^-$
 levels correctly reproduced by jj44b interaction. The negative levels predicted by $fpg$
 interaction is higher in energy. In contrast to experimentally observed $13/2^-$ doublet,
 the shell model predicted these levels more than 300 keV separation between each other.
For the $7/2_1^+$ level configuration is $\nu(g_{9/2}^{-3})$ with probability $\sim 13\%$ (JUN45)
and $\sim 18\%$ (jj44b), respectively. The calculated occupancies for the ground state
for neutron $g_{9/2}$ orbital
is 6.57 (JUN45) and 6.97 (jj44b). The JUN45 interaction predicts better results
for excitation energies in comparison to the jj44b and $fpg$ interactions.

\newpage
\begin{figure}
\includegraphics[width=14.4cm]{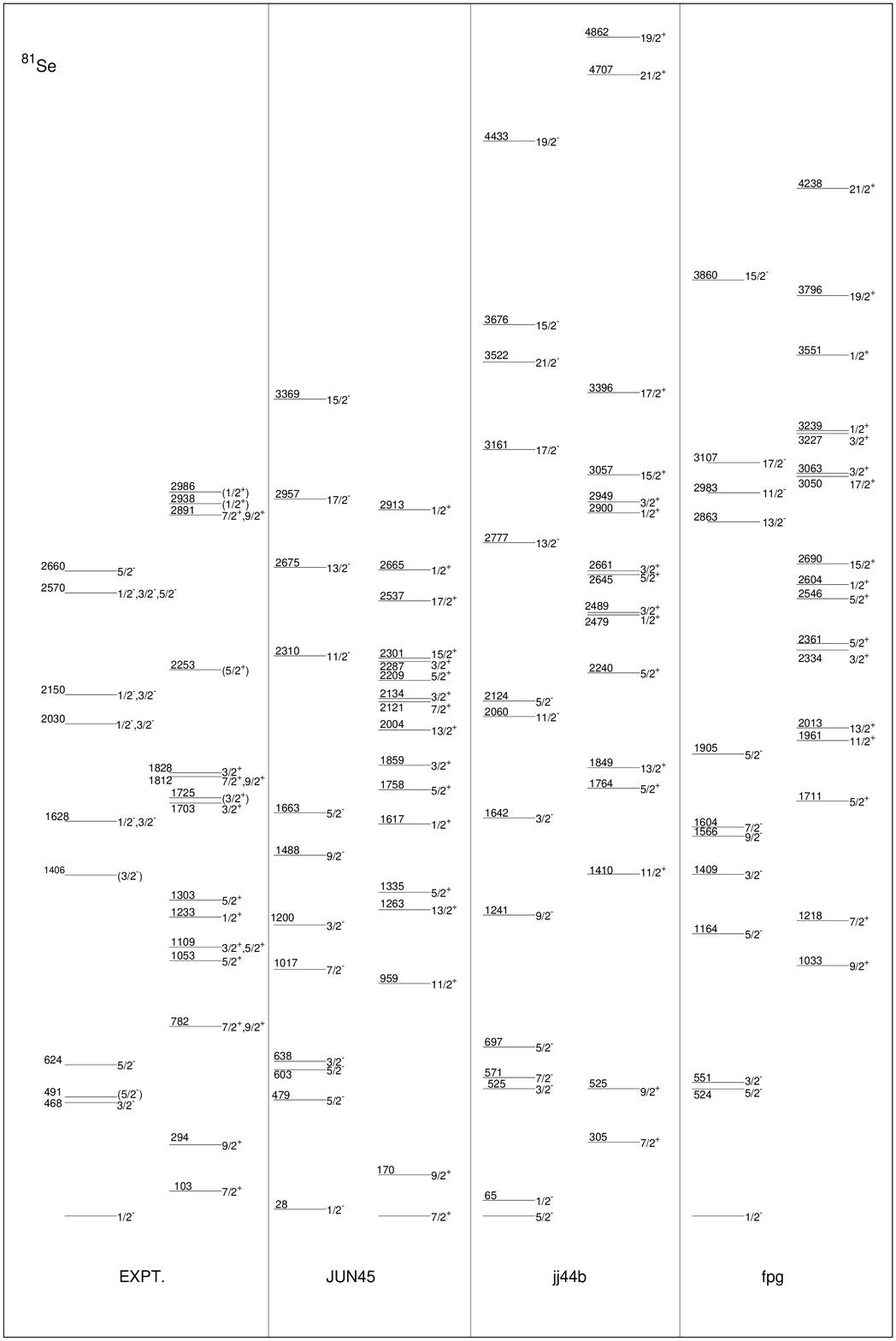}
\caption{
 The same as in Fig. 2, but for $^{81}$Se. }
\label{f_se81}
\end{figure}
\newpage
\begin{figure}
\includegraphics[width=14.4cm]{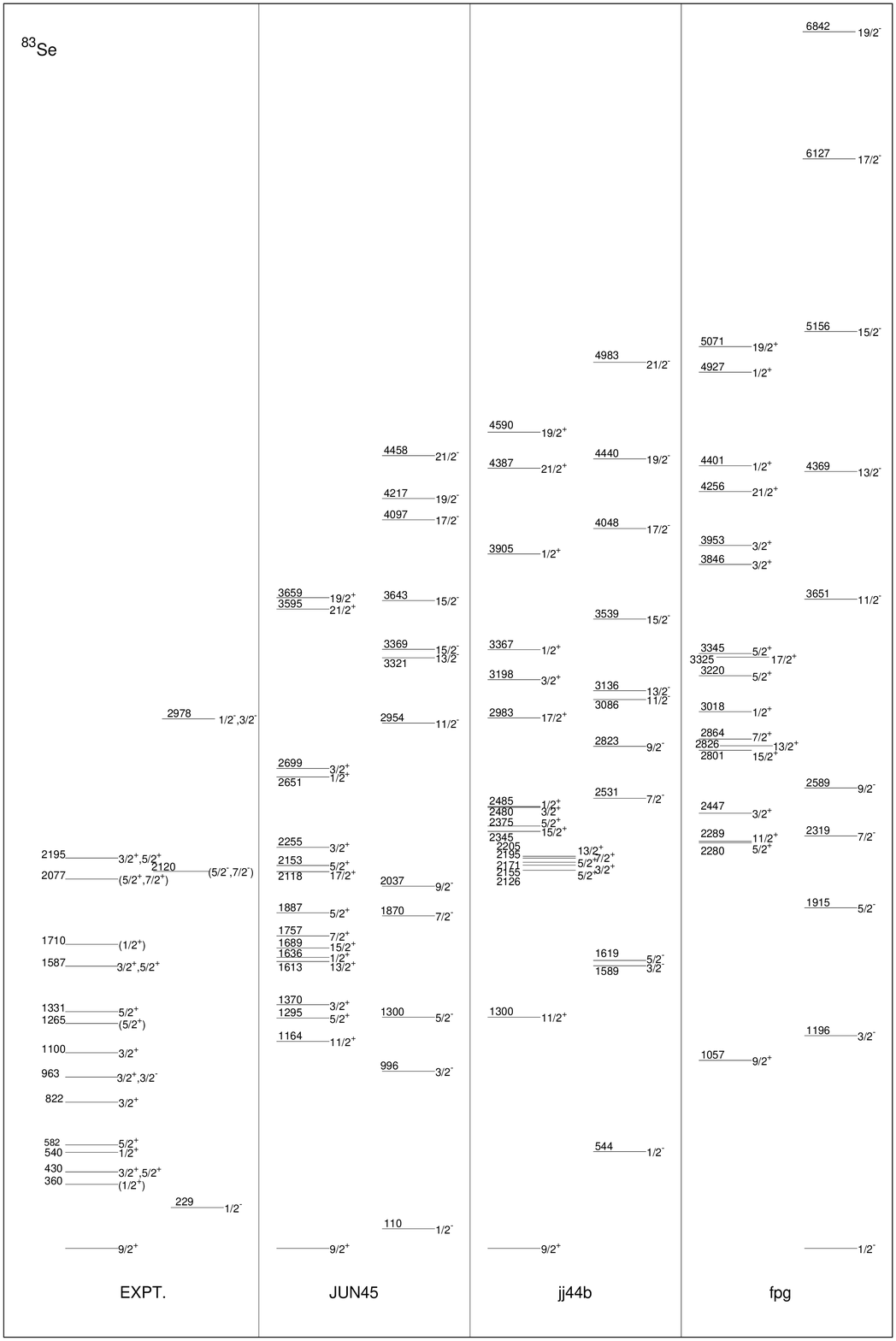}
\caption{
 The same as in Fig. 2, but for $^{83}$Se.}
\label{f_se83}
\end{figure}
\subsection{$^{81}$Se}

The comparison of calculated and experimental positive and negative-parity energy levels of $^{81}$Se
is given in Figure~\ref{f_se81}. Only $fpg$ interaction is able to correctly reproduce ground
state spin and parity. In jj44b interaction the first negative-parity
level is different from the experimental ones while JUN45 predicted correctly this level.
First two energy levels $7/2^+$ and $9/2^+$
for the positive-parity by both the interactions are in sequence with the experimental data 
but observed experimental difference of 191 keV predicted by JUN45 at 170 keV and with jj44b
at 220 keV, respectively. For the ground state i.e. $1/2_1^-$ level configuration is $\nu(p_{1/2}^{-1})$
with probability $\sim 36\%$ (JUN45)
and $\sim 30\%$ (jj44b), respectively. The calculated occupancies for ground state
for neutron $g_{9/2}$ orbital is 8.15 (JUN45) and 8.12 (jj44b). 
For the $7/2_1^+$ level configuration is $\nu(g_{9/2}^{-3})$ with probability $\sim 35\%$ (JUN45)
and $\sim 31\%$ (jj44b), respectively. The calculated occupancies for $7/2_1^+$
for neutron $g_{9/2}$ orbital is 7.20 (JUN45) and 7.22 (jj44b). These results demonstrate
importance of neutron $g_{9/2}$ orbital for the ground state.
The overall results of JUN45 is in good agreement with experimental data.

\subsection{$^{83}$Se}

For this isotope the experimental data is very sparse.
Figure~\ref{f_se83}, shows the calculated and experimental levels of $^{83}$Se using JUN45,
jj44b and $fpg$ interactions. The JUN45 and jj44b correctly reproducing ground state as $9/2^+$,
while $fpg$ interaction predicting  $1/2^-$ as a ground state. The energy level $1/2^+$ is at
1636 keV in JUN45 and at 2485 keV in jj44b while in experiment it is at 360 keV. 
The ${f_{5/2}pg_{9/2}}$ space 
based interactions predicting $11/2^+$ as first excited positive parity level, while
there is no experimental result for this level. The configuration for g.s. as $9/2^+$ is
$\nu(g_{9/2}^{-1})$ with probability of $\sim 53\%$ (JUN45)
, $\sim 46\%$ (jj44b) and $\sim 50\%$ ($fpg$), respectively. With both model spaces the 
first  $1/2^-$ having configuration
$\nu(p_{1/2}^{-1})$, with the maximum probability for this level is $\sim 70\%$ for $fpg$ interaction.
In comparison of experimental results of low-lying levels and corresponding high values of theoretical
results it is clear that neutrons excitation across $N=50$ shell is important. 
The calculated results are
high in energy which reflects that as we approach towards $N=50$, the calculation should include $d_{5/2}$
orbital in the model space. Because $f_{5/2}pg_{9/2}$ space is not enough.

In the work of Yoshinaga et al, ~\cite{yosinaga} for  $^{79}$Se the predicted $1/2^-$ level lies higher in
energy compared to experimental data, while with jj44b it is close to experimental data
with a difference of only 29 keV. In case of $^{83}$Se, the results predicted by JUN45 is better
than previous result.~\cite{yosinaga}

The occupancies of proton and neutron orbitals for $^{79,81,83}$Se isotopes, for the ground state 
and $1/2^-$ state are shown in Table~\ref{t_o}. Also in Figure 6, we shown
the proton and the neutron occupation numbers of the different orbitals. The proton occupancies for
$d_{5/2}$ orbital increase smoothly at the expense of the $p_{3/2}$
as we move from $^{79}$Se to $^{83}$Se. In case of $fpg$ interaction the occupancies of $f_{7/2}$
orbital is very dominant this reflects importance of inclusion of this orbital in the model space.
The neutron occupancies for
$g_{9/2}$ orbital increase smoothly at the expense of the $p_{3/2}$ and $f_{5/2}$ orbitals 
as we move from $^{79}$Se to $^{83}$Se. This reveals the importance of neutron excitation across
$N=40$.

\begin{figure}
\begin{center}
\includegraphics[scale=0.56]{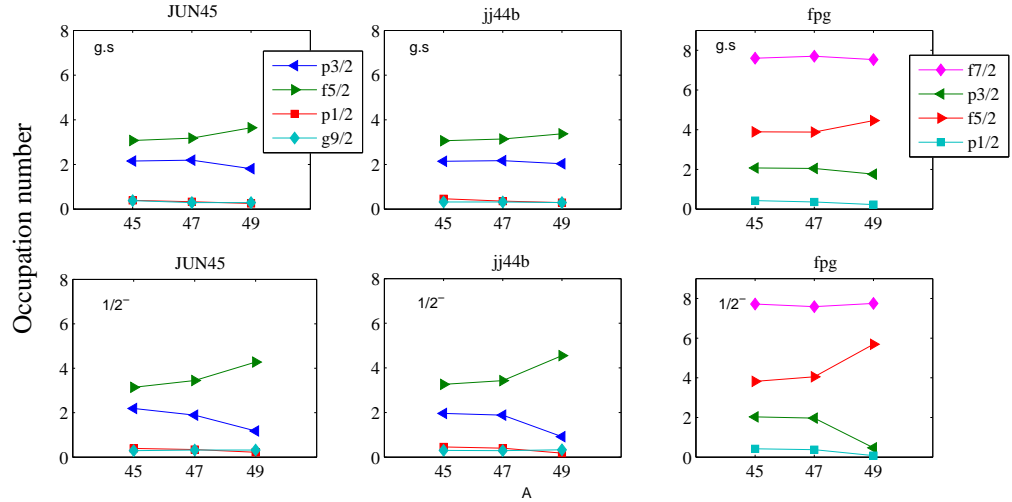}
\begin{center}
(a) Proton 
\end{center}
\includegraphics[scale=0.55]{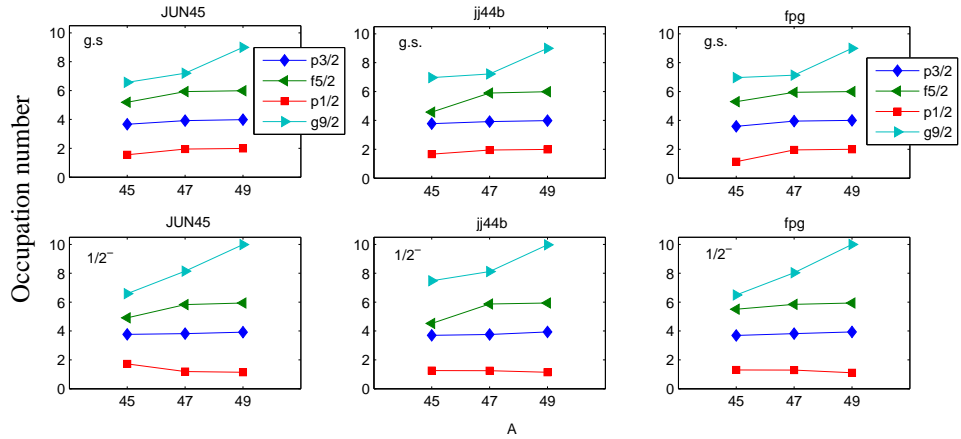}
\begin{center}
(b) Neutron 
\end{center}
%\epsfig{file=78Ge.eps, angle=-90, width=0.6\linewidth}
%\resizebox{90mm}{!}{\includegraphics{78Ge_1.eps}}
\caption{\label{Fig10}(Color online) Proton/Neutron occupation numbers of the JUN45 and jj44b  ($p_{3/2}$, $f_{5/2}$, $p_{1/2}$ and $g_{9/2}$ -shell orbits) and $fpg$ ($f_{7/2}$, $p_{3/2}$, $f_{5/2}$, $p_{1/2}$ -shell orbits) interactions-
 for two low-lying states in $^{79,81,83}$Se isotopes. Upper
panel for $7/2_1^+$ (g.s. in $^{79,81}$Se) and $9/2_1^+$ (g.s. in$^{83}$Se); lower panel for $1/2_1^-$ (in $^{79,81,83}$Se).
}
\end{center}
\label{f_82ge}
\end{figure}
% Occupancies
%check title
\begin{table}[h]
\caption{ Occupation of proton and neutron orbitals for $^{79,81,83}$Se isotopes in $f_{5/2}pg_{9/2}$ and $fpg_{9/2}$ spaces.}
\begin{center}
\resizebox{13.5cm}{!}{
%\begin{tabular}{|r|c|r||c|r|c|r|c||r|c|r|c|r|}
 \begin{tabular}{|r|c|r||c|r|c|r|c||c|r|c|r|}

\hline
%Interaction &Nucleus &$I$&$\pi0f_{7/2}$&$\pi1p_{3/2}$&$\pi1p_{1/2}$~~&$\pi0f_{5/2}$ ~~~~~~&$\pi0g_{9/2}$&$\nu0f_{7/2}$&$\nu1p_{3/2}$&$\nu1p_{1/2}$~~&$\nu0f_{5/2}$ ~~~~~~&$\nu0g_{9/2}$ \\
Interaction &Nucleus &$I$&$\pi0f_{7/2}$&$\pi1p_{3/2}$&$\pi0f_{5/2}$&$\pi1p_{1/2}$&$\pi0g_{9/2}$ &$\nu1p_{3/2}$&$\nu0f_{5/2}$~~&$\nu1p_{1/2}$&$\nu0g_{9/2}$ \\
\hline
  JUN45~~ &$^{79}$Se &$7/2_1^+$& &2.15 &3.07&0.39~~ &0.38 %& 
                                 &3.67 &5.19&1.55~~ &6.57 \\
          &    &$1/2_1^-$& &2.19 &3.14 &0.39~~  &0.29 %&
                                 &3.77  &4.91 &1.72~~&6.59 \\
\hline
 jj44b~~ &$^{79}$Se &$7/2_1^+$& &2.14 &3.06&0.46~~  &0.32 %&
                                &3.78 &4.58&1.66~~ &6.97 \\
          &    &$1/2_1^-$& &1.96 &3.26 &0.46~~ &0.30 %&
                             &3.70 &4.53 &1.26~~ &7.48 \\
\hline
$fpg$~~~~ &$^{79}$Se &$7/2_1^+$ &7.60&2.07  &3.89 &0.43~~ & \-- %0.00 %&8.00
                                &3.59  &5.30 &1.14~~ &6.97\\
                     &  & $1/2_1^-$&7.72&2.03 &3.82 &0.42~~ & \-- %0.00 %&8.00
                                &3.69  &5.51 &1.30~~ &6.50\\
\hline

\hline
  JUN45~~ &$^{81}$Se &$7/2_1^+$& &2.19 &3.18 &0.33~~  &0.29 %&
                                &3.92 &5.93 &1.95~~  &7.20 \\
                     &  &$1/2_1^-$& &1.89 &3.45&0.34~~   &0.32 %&
                                &3.82 &5.83&1.19~~   &8.15\\
\hline
 jj44b~~~~  &$^{81}$Se &$7/2_1^+$& &2.17 &3.14 &0.36~~ &0.32 %&
                                 &3.92  &5.90 &1.95~~ &7.22\\
                        &     &$1/2_1^-$& &1.88 &3.43&0.40~~ &0.29%&
                              &3.76  &5.87 &1.25~~ &8.12 \\
\hline
$fpg$~~~~ &$^{81}$Se &$7/2_1^+$& 7.70&2.05&3.88 &0.36~~ & \-- %0.00 %& 8.00
                               &3.95 &5.95 &1.96~~ &7.14 \\
                       &      &$1/2_1^-$ &7.59&1.97 &4.05&0.38~~ & \-- %0.00 %&8.00
                                &3.82 &5.85 &1.29~~ &8.04 \\
\hline

\hline
  JUN45~~ &$^{83}$Se &$9/2_1^+$& &1.81 &3.64 &0.25~~  &0.29 %&
                                 &3.99 &5.99 &1.99~~  &9.00 \\
                       &  &$1/2_1^-$& &1.18 &4.28&0.22~~   &0.32 %&
                                      &3.92 &5.94&1.14~~   &9.99\\
\hline
 jj44b~~~~  &$^{83}$Se &$9/2_1^+$& &2.03 &3.37 &0.29~~ &0.29 %&
                                   &3.99  &5.99 &1.99~~ &9.00\\
                       &     &$1/2_1^-$& &0.92 &4.55&0.18~~ &0.33%&
                                  &3.94  &5.94 &1.14~~ &9.97 \\
\hline
$fpg$~~~~ &$^{83}$Se &$9/2_1^+$& 7.53&1.76&4.46 &0.23~~ & \-- %0.00 %& 8.00
                               &4.00 &6.00 &2.00~~ &9.00\\
                    &      &$1/2_1^-$ &7.75&0.47 &5.69&0.08~~ & \-- %0.00 %&8.00
                               &3.94 &5.94 &1.11~~ &10.00 \\                                 
\hline\end{tabular}}
\label{t_o}
\end{center}
\end{table}

\begin{table}
\begin{center}
\caption{$B(E2)$ reduced transition strength in W.u. Effective charges
  $e_p=1.5$ $e_n=0.5$ were used.}
\vspace{2mm}
\begin{tabular}{ c  c  c  c  c  c  c  c}  \hline %\hline
             & $^{79}$Se & $^{81}$Se    & $^{83}$Se &  & $^{79}$Se & $^{81}$Se & $^{83}$Se\\ \hline
 BE($9/2_1^+ \rightarrow 7/2_1^+$) & & & & BE($9/2_1^+ \rightarrow 5/2_1^+$) & & & \\ \hline
Experiment  & N/A   & N/A    & N/A  & Experiment & N/A & N/A   & N/A\\ %\hline
JUN45       & 20.37 & 17.65  & 2.40  &JUN45     & 0.05 & 3.20  & 1.37\\ %\hline
jj44b       & 27.77 & 17.60  & 0.13  &jj44b    & 1.78  & 1.63  & 0.005\\ %\hline
$fpg$      & 32.03  &10.27   & 9.94  & $fpg$   & 12.13  &  8.09  & 7.84 \\  %\hline
\hline
\label{t_b}
\end{tabular}
\end{center}
\end{table}

\begin{table}
\begin{center}
\caption{ Electric quadrupole moments, $Q_s$ (in eb), the effective
charges $e_p$=1.5, $e_n$=0.5 are used in the calculation and Magnetic moments, $\mu$ (in $\mu_N$), for $g_{s}^{eff}$ = 0.7$g_{s}^{free}$. }
\hspace{-2mm}
\label{tab:table3}
%\resizebox{7.6cm}{4.5cm}{
\begin{tabular}{ c  c  c  c c  c  c  c } \hline
 & $^{79}$Se &$^{81}$Se&$^{83}$Se & & $^{79}$Se &$^{81}$Se&$^{83}$Se\\ \hline
 $Q$($7/2_1^+$) & & &  &  $\mu$($7/2_1^+$) & & &  \\ \hline
Experiment & +0.8 (2)  &N/A &N/A & Experiment & -1.018 (15) & N/A & N/A  \\ %\hline
JUN45 & +0.38 & +0.53 & +0.16 & JUN45 & -1.17 & -1.15 & -1.03 \\ %\hline
jj44b & +0.59 &+0.55 & +0.31 & jj44b & -1.29 &-1.13 & -1.16  \\ %\hline
$fpg$ &  +0.65 & +0.60  &  -0.07  &  $fpg$ & -1.20   & -1.24  & +2.47  \\ %\hline
       & & &  & & & & \\ \hline
$Q$($9/2_1^+$) &  & & &$\mu$($9/2_1^+$) & & &    \\ \hline
Experiment & N/A & N/A  & N/A  &Experiment & N/A & N/A  & N/A  \\ %\hline
JUN45 & +0.09 &+0.22 & +0.49 &JUN45 & -1.03 &-1.27 & -1.35  \\ %\hline
jj44b & +0.20 & +0.30 & +0.60 & jj44b & -1.37 & -1.23 & -1.45  \\ %\hline
$fpg$ & +0.25  & +0.49  & +0.61 & $fpg$ & -1.09  & -1.43  &   -1.35     \\ %\hline
       & & &  & & & &\\ \hline

$Q$($5/2_1^+$) &  & &  & $\mu$($5/2_1^+$) &  & &    \\ \hline
Experiment &N/A  &N/A &N/A & Experiment &N/A  &N/A &N/A  \\ %\hline
JUN45 & -0.002 &+0.01 &-0.01 & JUN45 & -0.93 &-1.54 &-1.72   \\ %\hline
jj44b & -0.11 &+0.28 &+0.14  & jj44b & -1.02 &-1.76 &-1.16  \\ %\hline
$fpg$ & -0.03  & +0.32  & +0.12  & $fpg$ & -1.15  & -1.63  & +0.65     \\ %\hline
       & & &   & & &  &\\ \hline
\label{t_q}     
\end{tabular}\end{center}
\end{table}

\section{\label{sec4}Electromagnetic properties}
The calculated $B(E2)$ transition probabilities for both model spaces are given in Table~\ref{t_b}. For this
effective charges $e_p$=1.5, $e_n$=0.5 
are used in the calculation.  The results of quadrupole moments and magnetic moments 
for the three different interactions are
shown in Table~\ref{t_q}. The experimental data for quadrupole moments for $^{79}$Se
show very good agreement with results of $fpg$ interaction.
This shows the importance of inclusion of $\pi f_{7/2}$ orbital in the model space,
which was proposed in.~\cite{Cheal10} The experimental data for
magnetic moments are available only for $^{79}$Se. In Table~\ref{t_q},
we compare calculated values of magnetic moments with the experimental data
for $^{79}$Se. We also have presented predicted values of magnetic moments by
the three interactions for the remained isotopes considered.
It is seen from Table~\ref{t_q} that for the $^{79}$Se calculated values of
magnetic moments are in better agreement with the experiment when JUN45 interaction
is used.  From Table we can see that the transition rates are strongly enhanced
while the quadrupole moments are of the order of the single-particle ones. The large
$B(E2)$ values are due to strong dynamical collectivity as these nuclei of this
region lose the magicity properties.

\section{\label{sec5}Conclusions}

We have reported shell model results for neutron-rich
odd Se isotopes for two spaces: full $f_{5/2}pg_{9/2}$ space
and $fpg_{9/2}$ space with $^{48}$Ca core. The following broad 
conclusions can be drawn: 

\begin{itemize}  

\item The energy levels,
$B(E2)$'s, quadrupole moments and magnetic moments
are in good agreement with the experimental data when
available.

\item  The $E2$ transitions, quadrupole moments and
magnetic moments analysis
show the importance of proton excitations across $Z=28$ shell for
 $fpg_{9/2}$ space. 

\item  Further theoretical development is needed by enlarging model
space by including $\nu d_{5/2}$ orbital to study simultaneously
proton and neutron excitations across $Z=28$  and $N=50$ shell,
respectively.

\end{itemize} 

Thanks are due to
E. Padilla - Rodal for useful discussions during this work.
All the shell-model calculations carried out at KanBalam
computational facility of DGCTIC-UNAM, Mexico. 
PCS would like to thank M.J. Ermamatov for useful discussions
about collectivity in this region. One of us VK acknowledges financial 
support from CSIR, India for his PhD work.

\end{document}